\documentclass[a4paper, 10 pt, conference]{ieeeconf}  % Comment this line out if you need a4paper

\IEEEoverridecommandlockouts                              % This command is only needed if 
\usepackage{epsfig} % for postscript graphics files
\usepackage{subcaption}
\usepackage{multirow}
\title{\LARGE \bf
Performing Stateful Logic Using Spin-Orbit Torque (SOT) MRAM
}

\author{Barak Hoffer and Shahar Kvatinsky% <-this % stops a space
\vspace{-5pt}
\thanks{*This work was partially supported by the European Research Council through the European Union's Horizon 2020 Research and Innovation Programe under Grant 757259,
and partially supported by the European Research Council through the European Union's Horizon Europe Research and Innovation Programe under Grant 101069336.}% <-this % stops a space
\thanks{Barak Hoffer and Shahar Kvatinsky are with the Andrew and Erna Viterbi Faculty of Electrical and Computer Engineering, Technion – Israel Institute of Technology, Haifa 3200003, Israel (e-mail: barakhoffer@campus.technion.ac.il; shahar@ee.technion.ac.il).}%
}

\begin{document}

\maketitle
\pagestyle{empty}
\pubid{\begin{minipage}{\textwidth}\ \\[12pt] \centering \copyright 2022 IEEE. Personal use of this material is permitted. Permission from IEEE must be obtained for all other uses, in any current or future media, including reprinting/republishing this material for advertising or promotional purposes, creating new collective works, for resale or redistribution to servers or lists, or reuse of any copyrighted component of this work in other works.\end{minipage}}

%%%%%%%%%%%%%%%%%%%%%%%%%%%%%%%%%%%%%%%%%%%%%%%%%%%%%%%%%%%%%%%%%%%%%%%%%%%%%%%%
\begin{abstract}

Stateful logic is a promising processing-in-memory (PIM) paradigm to perform logic operations using emerging nonvolatile memory cells.
While most stateful logic circuits to date focused on technologies such as resistive RAM, we propose two approaches to designing stateful logic using spin-orbit torque (SOT) MRAM.
The first approach utilizes the separation of read and write paths in SOT devices to perform logic operations. In contrast to previous work, our method utilizes a standard memory structure, and each row can be used as input or output. The second approach uses voltage-gated SOT switching to allow stateful logic in denser memory arrays. We present array structures to support the two approaches and evaluate their functionality using SPICE simulations in the presence of process variation and device mismatch.

\end{abstract}

%%%%%%%%%%%%%%%%%%%%%%%%%%%%%%%%%%%%%%%%%%%%%%%%%%%%%%%%%%%%%%%%%%%%%%%%%%%%%%%%
\section{Introduction}
\pubidadjcol
As computer applications become more and more data-centric, the performance and power bottlenecks caused by the movement of data between the processor and memory become dominant.
Processing-in-memory (PIM) addresses these issues by adding computation capabilities to the memory. Stateful logic is a PIM technique that uses the unique electrical properties of emerging memristive technologies to implement logic gates directly, using the memory cells, without reading the data outside the memory array. Each memristive technology has its own switching mechanism, benefits and limitations; therefore, stateful logic has been evaluated and demonstrated using different technologies such as RRAM~\cite{Borghetti2010, Kvatinsky2014, Hoffer2020}, PCM~\cite{Hoffer2022}, STT-MRAM~\cite{Patil2010,Zabihi2019,Louis2019}, and recently, SOT-MRAM~\cite{Resch2020}.

Magnetoresistive random access memory (MRAM) is considered a promising memory technology in terms of speed, power consumption, and endurance~\cite{Shiokawa2019,Shao2021}. STT-MRAM has shown several benefits such as nonvolatility, low static power, and compact bit-cell size, compared to conventional SRAM-based memory. Nevertheless, current STT-MRAM technology is limited by read disturb and endurance issues~\cite{Shao2021}. Alternatively, SOT-MRAM offers lower power consumption, higher switching speed and endurance compared to STT-MRAM. In such devices, their three-terminal cell structure separates the write and read paths, thereby minimizing the risk of device breakdown and increasing device endurance.

Stateful logic has been been examined in the context of SOT-MRAM devices~\cite{Resch2020,Shreya2021}, targeting applications such as deep-learning inference~\cite{Kim2022}. To date, however, studies suggested using array structures with dedicated locations for inputs and outputs. This limits array density optimization and mapping of gates to the array. Furthermore, as SOT switching is still being researched and improved, it is unclear how close these methods are to actual realization, considering the current limitations of SOT-MRAM devices.

In this paper, we propose two techniques based on SOT-MRAM array memory structures to support stateful logic. Each technique relies on a different device switching mechanism. We describe the benefits and limitations of using each array structure and evaluate both methods using simulations in the presence of process variation and device mismatch.

\begin{figure}[tpb]
  \centering
  \begin{subfigure}[t]{.45\linewidth}
    \centering
    \includegraphics[width=\linewidth]{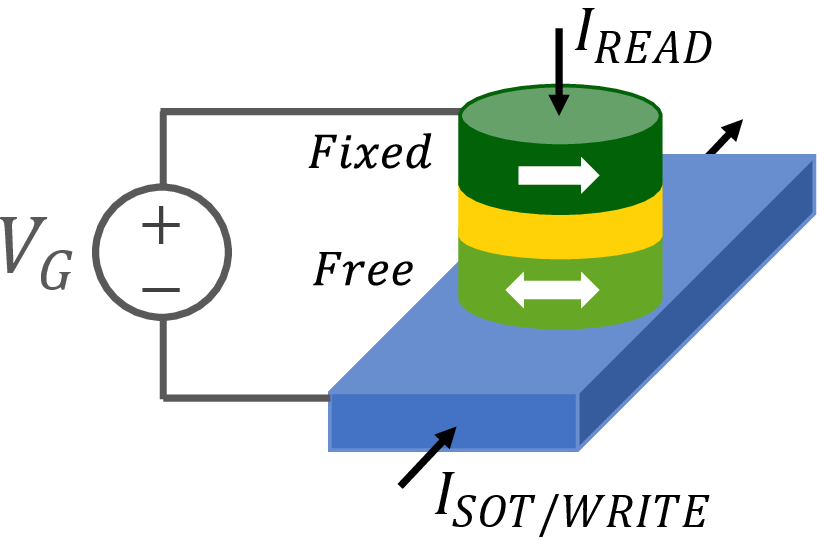}
    \caption{}
  \end{subfigure}
  \begin{subfigure}[t]{.45\linewidth}
    \centering
    \includegraphics[width=\linewidth]{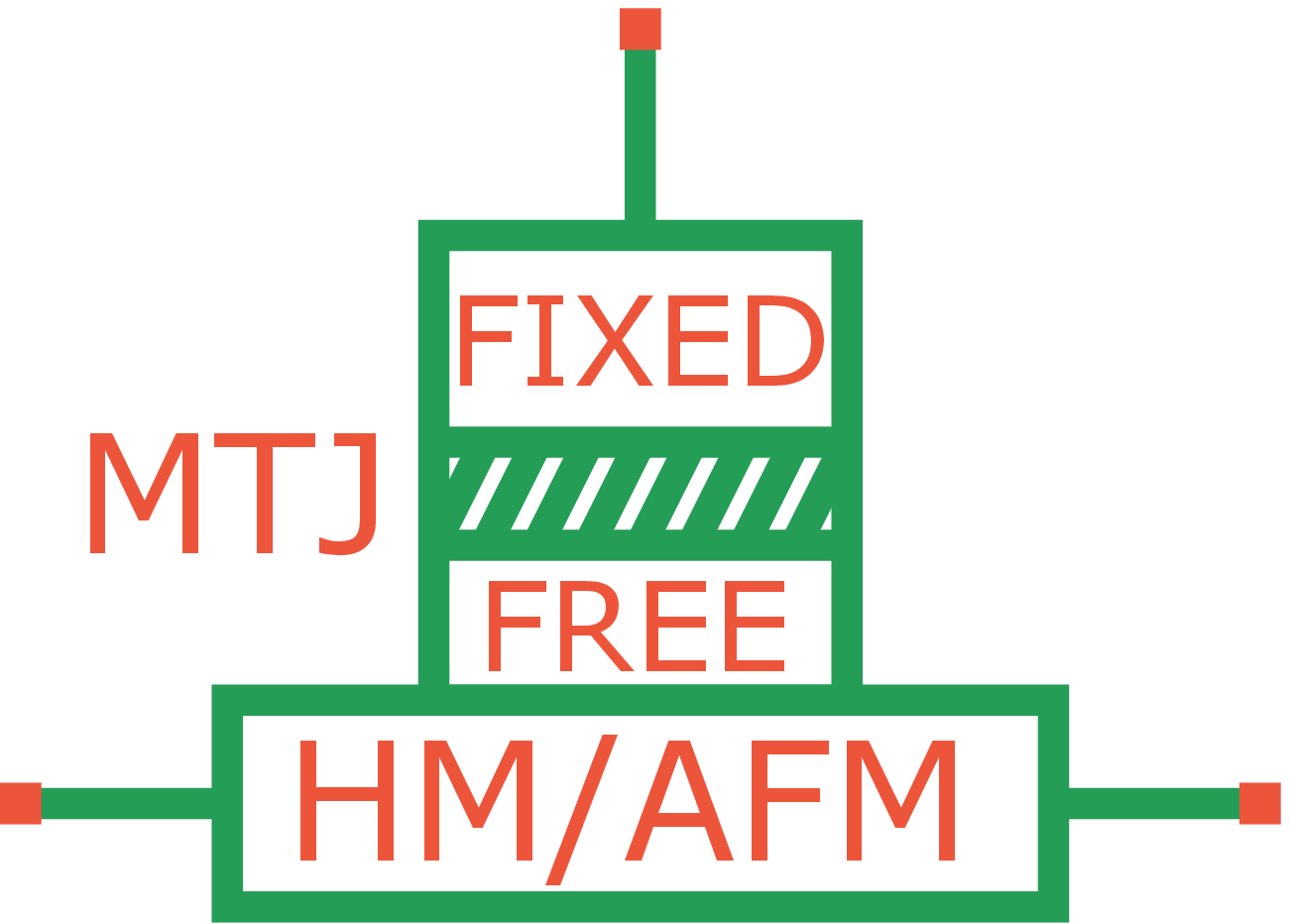}
    \caption{}
  \end{subfigure}
  \caption{SOT-MRAM cell (a) structure and (b) symbol.}
  \label{fig:device}
\end{figure}

\section{SOT-MRAM Devices}
\begin{figure*}[thpb]
  \centering
  \begin{subfigure}[t]{.55\textwidth}
    \centering
    \includegraphics[width=\textwidth]{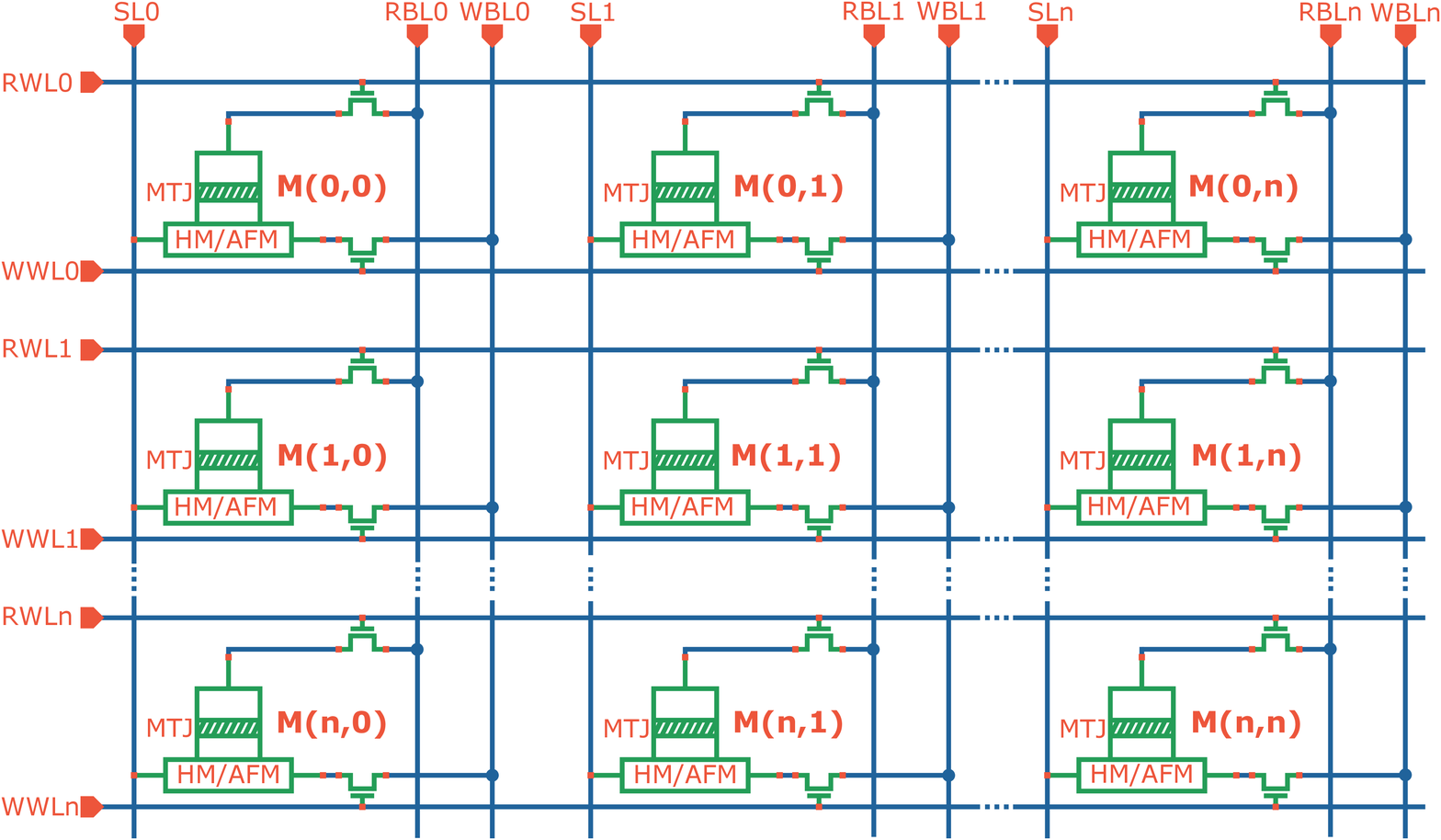}
    \caption{}
    \label{fig:std-array}
  \end{subfigure}
  \hfill
  \begin{subfigure}[t]{.35\textwidth}
    \centering
    \includegraphics[width=\textwidth]{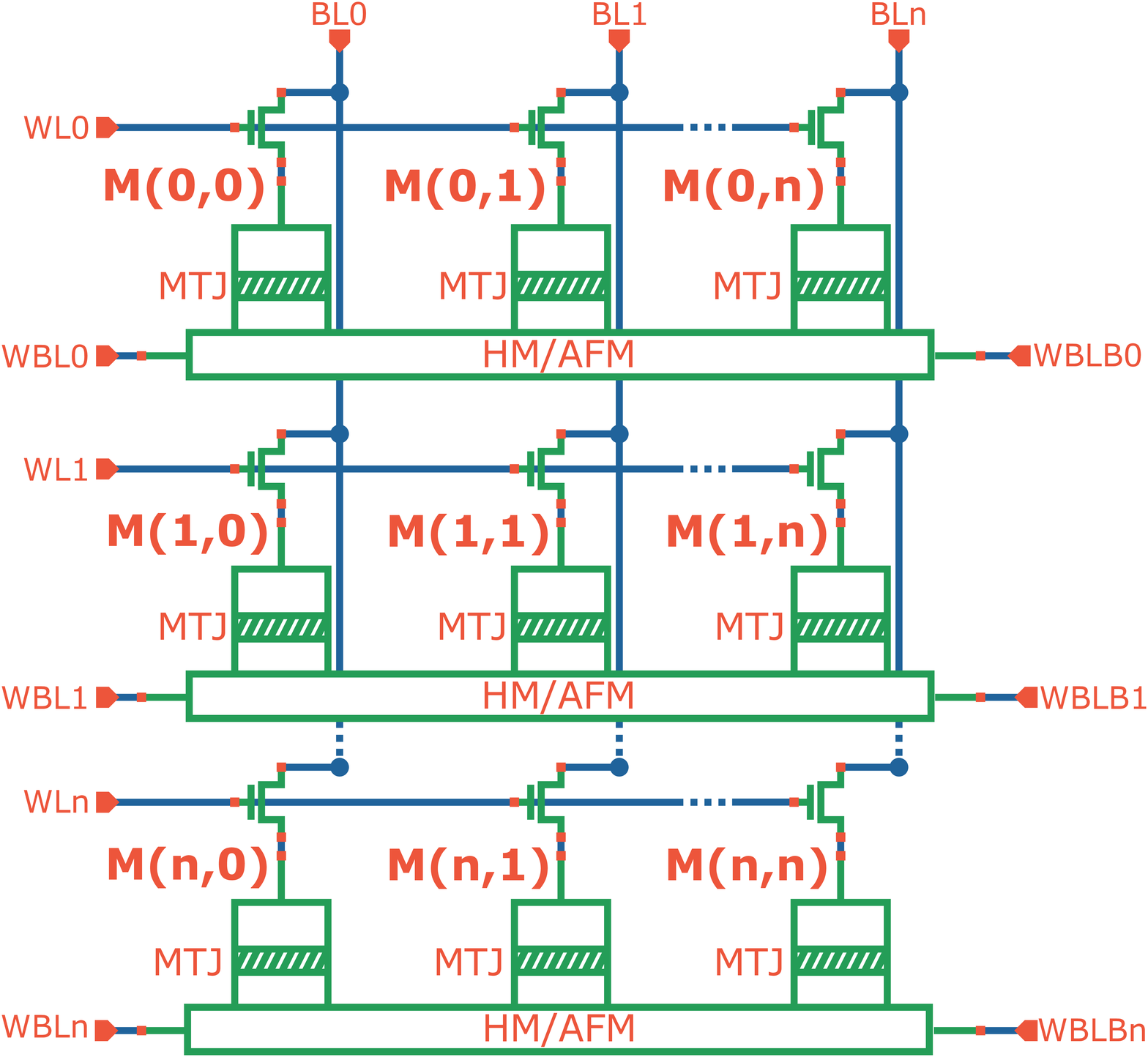}
    \caption{}
    \label{fig:vg-array}
  \end{subfigure}
  \caption{SOT-MRAM array structures supporting stateful logic. (a) A 2T-1R n$\times$n array. (b) A VGSOT-MRAM n$\times$n array. The voltage across the MTJ can lower the SOT current required for switching.}
  \label{fig:array}
\end{figure*}
\pubidadjcol
MRAM is constructed using magnetic tunnel junction (MTJ) devices, which comprise a fixed polarization layer and a free-layer, separated by an insulating layer. Data storage in MTJs is based on changing the magnetization of the free-layer. The electrical resistance of the device is determined by the relative magnetization of both layers. The resistance is lower when the magnetization of both layers is parallel ($R_{P}$), compared to the anti-parallel state ($R_{AP}$). Switching the free-layer can be achieved using two well-established mechanisms, spin-torque-transfer (STT)~\cite{Apalkov2013} or spin-orbit-transfer (SOT)~\cite{Garello2018}.

In STT-MRAM, passing a current through the MTJ creates a spin-polarized current that changes the magnetic polarity of the free-layer when it reaches a critical current.
For SOT-MRAM, a charge-current passing through an underlying heavy-metal (HM) or antiferromagnetic (AFM) layer injects a spin-current in the free-layer lying on top. A typical cell structure of an SOT-MRAM is shown in Fig.~\ref{fig:device}. The SOT switching mechanism decouples the read and write operations, allowing separate optimization of the read and write paths, unlike STT-MRAM. To lower the read currents, the MTJ can be fabricated with a high resistance-area product ($R\cdot A$). For write operations, the low resistance of the HM/AFM channel allows an efficient magnetization change, enabling a fast and energy-efficient write operation compared to STT-MRAM. To control the read/write operations, a transistor is added for each path. This constructs a two-transistor, one-MTJ (2T-1R) array, as depicted in Fig.~\ref{fig:std-array}. 

Another switching mechanism for SOT-MRAM is based on the voltage-gated SOT (VGSOT) effect~\cite{Wu2021}, which utilizes voltage-controlled magnetic anisotropy (VCMA) to assist the SOT. Based on the VCMA effect, the energy barrier for switching the MTJ between parallel and anti-parallel states is reduced when a positive bias voltage is applied to the oxide layer of the MTJ. Therefore, using the VCMA effect in this array structure reduces the critical SOT current, thereby enhancing the switching reliability and reducing the switching energy dissipation. Moreover, in VG-based SOT, the underlying layer can be shared between MTJ cells, and the switching is enabled only on cells where we apply the VCMA effect. As shown in Fig.~\ref{fig:vg-array}, the VCMA effect enables denser arrays, with bit-cell area of approximately half compared to standard SOT-MRAM~\cite{Wu2021}, since the SOT channel for each row is shared and requires only a single transistor per cell (1T-1R).
\begin{figure}[thpb]
  \centering
  \begin{subfigure}[t]{0.43\linewidth}
    \centering
    \includegraphics[width=\linewidth]{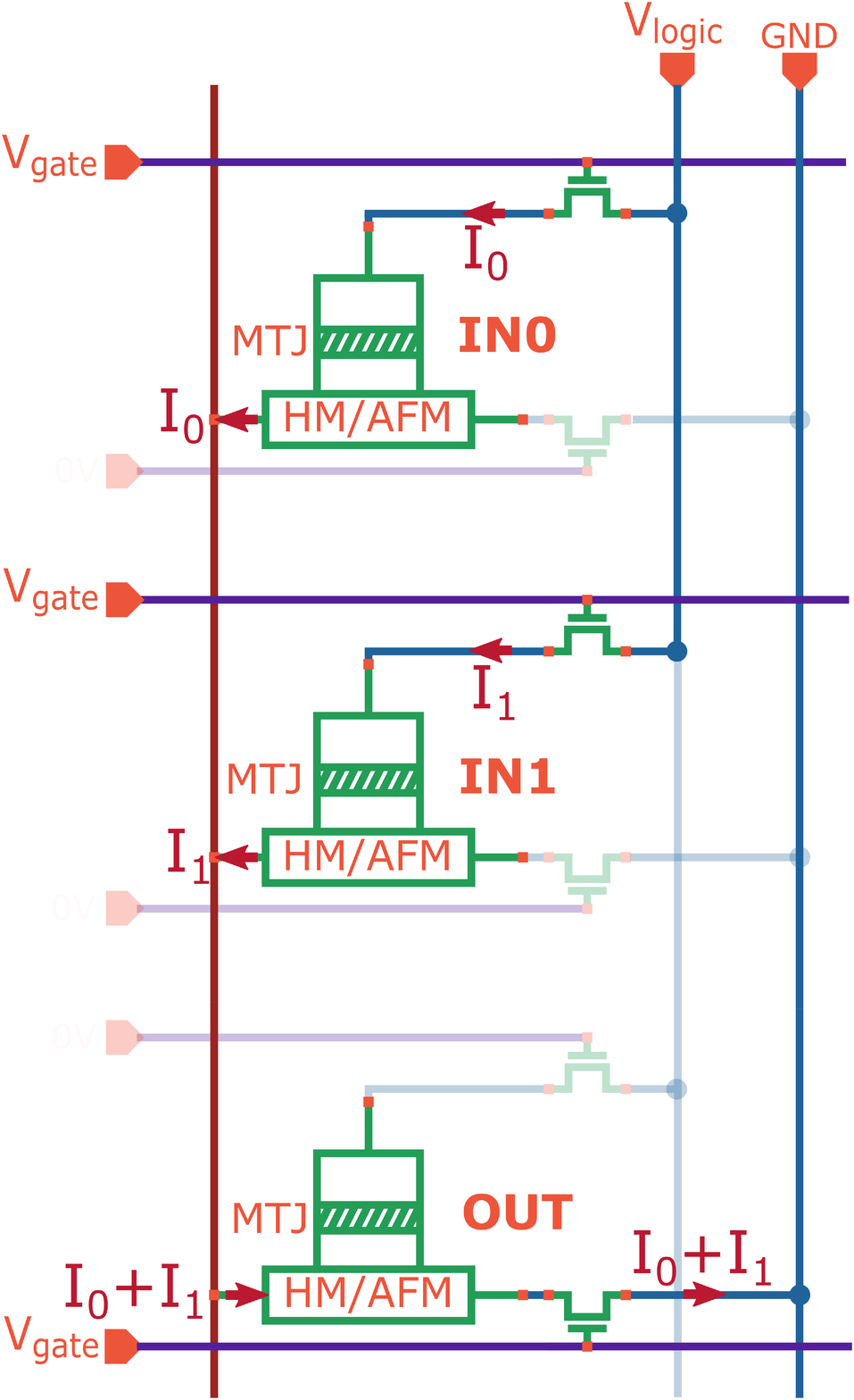}
    \caption{}
    \label{fig:std-stateful}
  \end{subfigure}
  \hfill
  \begin{subfigure}[t]{0.4\linewidth}
    \centering
    \includegraphics[width=\linewidth]{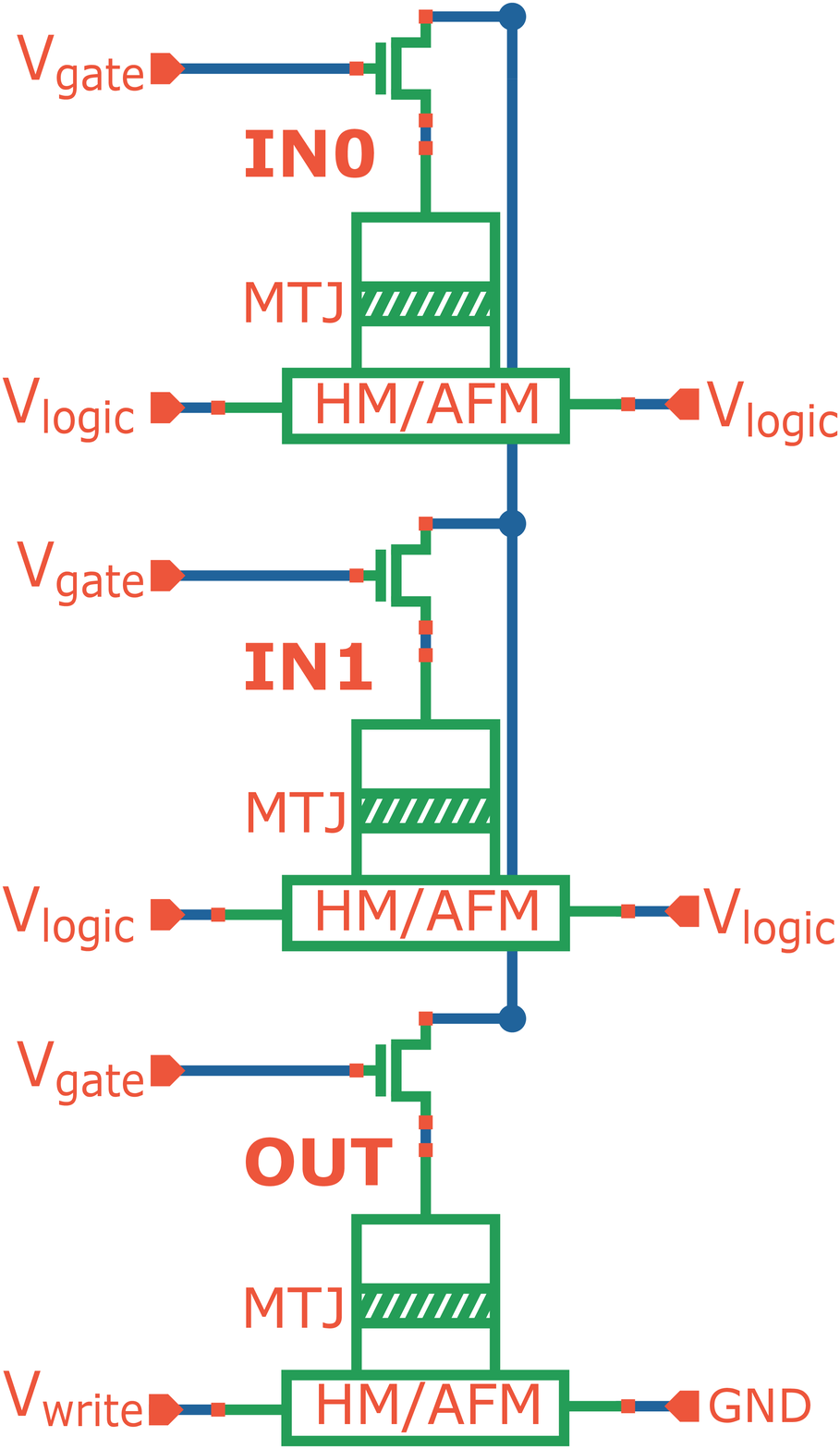}
    \caption{}
    \label{fig:vg-stateful}
  \end{subfigure}
  \caption{Stateful logic gates across rows in SOT-MRAM arrays. (a) 2T-1R SOT-MRAM array-- the read current from the input cells conditionally switches the output cell. (b) VGSOT-MRAM array-- the voltage across the shared BL depends on the state of the inputs, and conditionally lowers the SOT switching current.}
  \label{fig:stateful}
  \vspace{-5pt}
\end{figure}
\section{Statful Logic Using SOT-MRAM}

A stateful logic gate consists of multiple memory cells serving as inputs and output of the gate, where their resistance represents the logical states. Usually, the operation is based on a conditional switching of the output due to a voltage divider between the input and output devices~\cite{Kvatinsky2014,Hoffer2020,Hoffer2022}. In SOT devices, since the read and write paths are separated, a new scheme is required. Previous works suggested modifying the array structure to support stateful logic in SOT-MRAM by using the read current of the input cell to conditionally switch the output. However, the array structures in those studies were specifically arranged for the stateful operations, with separate array sections for inputs and outputs~\cite{Kim2022}. Other works suggested structures with specific line connections for even/odd columns, with the column parity of inputs and output forced to be opposite~\cite{Resch2020,Shreya2021}. These structures are harder to optimize in terms of density and require more complex control schemes for memory or logic. Additionally, since these structures are not standard memory, gate mapping might be less efficient. Our proposed methods use standard memory array structures where logic operations can be performed just by operating multiple rows, while any row can be used as input or output.

\subsection{2T-1R Array}
To support stateful logic operations in a 2T-1R array, we use an already demonstrated SOT array structure~\cite{Chen2021}, shown in Fig.~\ref{fig:std-array}, where each column of cells shares the select-line (SL), read bit-line (RBL) and write bit-line (WBL). This standard memory array structure does not require additional connections to operate the gates, and its orderly structure allows density optimizations and ease of control. Operation of a specific row is controlled by the write word-line (WWL) and the read word-line (RWL). As depicted in Fig.~\ref{fig:std-stateful}, operating a stateful logic operation in such an array is achieved by opening the read transistors of one or more input cells and the write transistor of the output cell. When a voltage is applied to the RBL, while grounding the WBL and keeping the SL floating, the current passing through the HM/AFM channel depends on the resistance state of the input cells.

To perform a NOR gate, we initialize the output cell to $R_{P}$ (`1') and apply 1.1V across the RBL. The gate voltage of the transistors is also at 1.1V. When at least one of the inputs is $R_{P}$ (`1'), the current passing in the SOT channel of the output is sufficient to switch the output to $R_{AP}$ (`0'). Additional gates can be constructed in the same manner. For example, by lowering the RBL voltage, a NAND gate can be designed since the output will switch only when both inputs are $R_{P}$ (`1'). Furthermore, by initializing the output cell to $R_{AP}$ (`0') and reversing the polarity of the voltage applied to the RBL, the OR and AND gates can be operated.

This simple mechanism has a few limitations and requirements with respect to the SOT-MRAM characteristics. First, the maximum accumulated read current needs to be higher than the critical current for SOT switching. This condition limits the maximum $R\cdot A$ of the MTJ, since the accumulated read current of the inputs is the output SOT write current. Currently, the SOT critical current in state-of-the-art devices is still in ranges higher than 100uA~\cite{Gupta2020}. Since MTJs are operated with voltages of a few volts, to generate such high currents using the read current, the resistances of the MTJ should be in the range of kiloohms, meaning $R\cdot A < 20 ~\Omega \cdot \mu m^2$. The requirement for high current also affects other factors of the MTJ, such as its size, since the current density in the MTJ during the operation should be lower than the critical current for STT switching, to prevent read disturbs in the inputs. Second, the tunnel magnetoresistance (TMR), which is the relative resistance change of the device between $R_{P}$ and $R_{AP}$, needs to be sufficiently high to create a margin between the read current of the switching and no-switching input cases, as the actual current will vary due to process and device variation. A smaller margin causes more computation errors. Currently, the TMR of MTJs is 100\%--250\%, which means a small margin. Previous works also suggested additional gates with more than two inputs~\cite{Shreya2021}, \textit{e.g.}, MAJ3. While this is possible in our suggested method as well, additional inputs will cause even smaller margins between states---which means the gate is more prone to errors.

\subsection{VGSOT-MRAM Array}

In VGSOT-MRAM arrays, we propose using voltage gating to execute stateful logic gates, as shown in Fig.~\ref{fig:vg-stateful}. Two rows serve as the input and a third row as the output, the gate voltage is applied to the WBL and WBLB of the inputs and the WBL and WBLB of the output are grounded. Then, for each column where the BL is kept floating, the voltage across the BL is defined by the voltage divider of the inputs and output. A higher voltage is induced across the output when the inputs are `1' ($R_{P}$). This voltage is also the gate-voltage of the output cell, reducing the critical SOT current. Now, when passing this lower SOT current across the output cell, it will switch the output device only when the BL voltage is high enough. In this type of stateful logic circuit, the $R\cdot A$ of the inputs is unbounded, since the current passing through the inputs does not induce the output switching directly. The $R\cdot A$ product, however, might limit the time required to charge the BL, so it will affect the latency of the gate.

For a stateful NOR gate in VGSOT arrays, we initialize the output cell to $R_{P}$ (`1'), apply 1.5V across the WBL and WBLB of the inputs and ground the WBL and WBLB of the output. When at least one of the inputs is $R_{P}$ (`1'), the voltage across the BL is sufficiently high to switch the output when we pass a current of $I_{SOT}=60\mu A$ from the WBL to the WBLB of the output. Other gates can be designed in a similar manner, for example, by lowering the WBL and WBLB voltages of the inputs we get a NAND gate and by initializing the output cell to $R_{AP}$ (`0') and reversing the polarity of the voltage applied to the WBL of the output, the OR and AND gates can be operated.
\section{Evaluation}
\begin{table}[t]
  \vspace{10pt}
  \caption{SOT-MRAM Model Parameters}
  \vspace{-10pt}
  \label{tab:parameters}
  \begin{center}
    \def\arraystretch{1.2}
    \begin{tabular}{|l|l|p{0.33\linewidth}|}
      \hline
      \textbf{Parameter} & \textbf{Description}     & \textbf{Value}                       \\ \hline \hline
      $D$                & MTJ diameter          & 50 nm                                \\ \hline
      $A_{MTJ}$          & MTJ surface area      & $\pi \cdot D^2 / 4$                  \\ \hline
      $t_f$              & Free layer thickness     & 1.1 nm                               \\ \hline
      $t_{ox}$           & MgO thickness            & 1.4 nm                               \\ \hline
      $M_s$              & Saturation magnetization & $6.25\times 10^5$ A/m                \\ \hline
      $K_i(0)$           & Interfacial PMA at 0V    & $3.2\times 10^{-4} J/m^2$            \\ \hline
      $\alpha$           & Gilbert damping factor   & 0.05                                 \\ \hline
      P                  & Spin polarization        & 0.58                                 \\ \hline
      $R\cdot A$         & Resistance-area product  & $10~\Omega \cdot \mu m^2~$ (2T-1R)  $650~\Omega \cdot \mu m^2$ ~(VGSOT)        \\ \hline
      TMR(0)             & TMR ratio at 0V             & 100\%                                \\ \hline
      $\beta$            & VCMA coefficient         & 60 fJ/V$\cdot$m                      \\ \hline
      $\theta_{SH}$      & Spin Hall angle          & 0.25                                 \\ \hline
      $H_{EX}$           & Exchange bias            & -50 Oe                               \\ \hline
      $L$                & SOT channel length       & 60 nm                                \\ \hline
      $W$                & SOT channel width        & 50 nm                                \\ \hline
      $T$                & SOT channel thickness    & 3 nm                                 \\ \hline
      $\rho_{SOT}$       & SOT channel resistivity  & $2.78 \times 10^{-6}~\Omega \cdot m$ \\ \hline
    \end{tabular}
  \end{center}
\end{table}

\begin{table}[t]
  \caption{2T-1R Array NOR Logic MC Results}
  \vspace{-10pt}
  \label{tab:std-success}
  \begin{center}
    \normalsize
    \def\arraystretch{1.3}
    \begin{tabular}{|c|c|c|c|c|}
      \hline
      \multirow{2}{*}{\textbf{IN1}} & \multirow{2}{*}{\textbf{IN0}} & \multirow{2}{*}{\textbf{OUT}} & \multicolumn{2}{c|}{\textbf{Success Rate}} \\ \cline{4-5}
      & & & 2T-1R & VGSOT \\ \hline \hline
      `0'          & `0'          & `1'          & 87.4\%    &    100\%        \\ \hline
      `0'          & `1'          & `0'          & 100\%     &     90\%          \\ \hline
      `1'          & `0'          & `0'          & 100\%     &    90\%        \\ \hline
      `1'          & `1'          & `0'          & 100\%     &    99.8\%      \\ \hline
    \end{tabular}
  \end{center}
   \vspace{-15pt}
\end{table}

\begin{figure}[tpb]
  \centering
    \includegraphics[width=\linewidth]{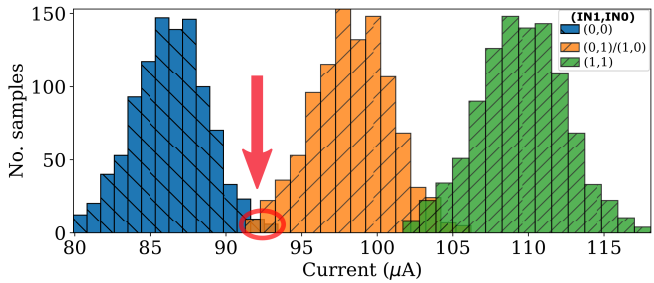}
  \caption{SOT current for the output cell during stateful NOR in a 2T-1R SOT-MRAM array, obtained by MC simulations (1000 samples). The red marking shows an overlap: some samples of the (0,0) input case have the same current as the (0,1)/(1,0) case, which causes unintentional switching.}
  \label{fig:currents}
 \vspace{-5pt}
\end{figure}

We evaluated the gates using SPICE simulations in Cadence Spectre, with GlobalFoundries 22FDX PDK to model the transistors, including process variation and device mismatch. To model the SOT-MRAM device, a Verilog-A model was used~\cite{Zhang2020}. The model captures STT, SOT and voltage-gated based switching, and considers thermal noise. The device parameters are listed in Table \ref{tab:parameters}, with a 3\% standard deviation of $t_{ox}, t_f$ and TMR.
The pulse width used to evaluate each operation is 2ns, and the maximum energy required for the NOR gate in 2T-1R and VGSOT arrays is 243$\mathrm{fJ}$ and 86$\mathrm{fJ}$, respectively.
Monte-Carlo (MC) simulation results for a NOR gate in each structure are given in Table \ref{tab:std-success}. For both arrays, each input was evaluated for 1000 samples. In the 2T-1R array scheme, the input IN1=IN2=`0' suffers from errors when the output is switched unintentionally due to variations and thermal noise. These errors can be explained by an overlap between current values for different inputs. Fig.~\ref{fig:currents} plots the SOT current for the output cell during the stateful NOR gate in a 2T-1R SOT-MRAM array, obtained by MC simulations. It shows an overlap in current values between some samples of the (0,0) input and the (0,1)/(1,0) input. Such an overlap might explain the unintentional switching. For the VGSOT array scheme, switching errors occur when variations of the gated-voltage cause the actual critical current to be higher than the current used. This results in switching not occurring.

\section{Conclusion}
In this paper, we explored stateful logic using SOT-MRAM. First, we showed a new method to operate stateful logic gates in existing SOT-MRAM memory arrays, instead of using specifically arranged arrays. Then, we showed a novel method for performing stateful logic in the denser and more energy efficient VGSOT-MRAM arrays.
We evaluated both methods in MC simulations, using existing SOT device arrays, considering process variation, device mismatch and thermal noise. The results show that the success rate of stateful logic using typical SOT-MRAM devices is limited by device characteristics and variations. For future devices, reducing the critical current required for SOT switching will allow 2T-1R stateful gates with lower currents, thus lower energy requirement. Improving SOT-MRAM devices' TMR will allow better separation between input cases, thus enhancing success rates for both methods.

\addtolength{\textheight}{-12cm}   % This command serves to balance the column lengths

%%%%%%%%%%%%%%%%%%%%%%%%%%%%%%%%%%%%%%%%%%%%%%%%%%%%%%%%%%%%%%%%%%%%%%%%%%%%%%%%

%%%%%%%%%%%%%%%%%%%%%%%%%%%%%%%%%%%%%%%%%%%%%%%%%%%%%%%%%%%%%%%%%%%%%%%%%%%%%%%%



% \section*{ACKNOWLEDGMENT}

% The preferred spelling of the word acknowledgment in America is without an e after the g. Avoid the stilted expression, One of us (R. B. G.) thanks . . .  Instead, try R. B. G. thanks. Put sponsor acknowledgments in the unnumbered footnote on the first page.

%%%%%%%%%%%%%%%%%%%%%%%%%%%%%%%%%%%%%%%%%%%%%%%%%%%%%%%%%%%%%%%%%%%%%%%%%%%%%%%%

\bibliographystyle{IEEEtran}
\bibliography{IEEEabrv,./ref}

\end{document}